\documentclass[conference]{IEEEtran}
\IEEEoverridecommandlockouts

\usepackage{cite}
\usepackage{amsmath,amssymb,amsfonts}
\usepackage{algorithmic}
\usepackage{graphicx}
\usepackage{textcomp}
\usepackage{xcolor}
\usepackage{booktabs}
\usepackage{multirow}
\usepackage{url}
\usepackage{microtype}
\usepackage{enumitem}

\def\BibTeX{{\rm B\kern-.05em{\sc i\kern-.025em b}\kern-.08em
    T\kern-.1667em\lower.7ex\hbox{E}\kern-.125emX}}

\setlist[enumerate]{nosep, leftmargin=*, topsep=1.5pt, itemsep=1.5pt}
\setlist[itemize]{nosep, leftmargin=*, topsep=1.5pt, itemsep=1.5pt}

\makeatletter
\g@addto@macro\normalsize{%
  \setlength\abovedisplayskip{3.0pt plus 1pt minus 1pt}%
  \setlength\belowdisplayskip{3.0pt plus 1pt minus 1pt}%
  \setlength\abovedisplayshortskip{1.0pt plus 0.5pt}%
  \setlength\belowdisplayshortskip{1.0pt plus 0.5pt}%
}
\makeatother

\let\oldbibliography\thebibliography
\renewcommand{\thebibliography}[1]{%
  \oldbibliography{#1}%
  \setlength{\itemsep}{0pt}%
  \setlength{\parskip}{0pt}%
}

\begin{document}

\title{Unlocking Heterogeneous Accelerator Flexibility in AI Data Centers: Sizing and Workload Allocation}

\author{
\IEEEauthorblockN{Chenghao Lyu, Ying Yu, Wangkun Xu, and Fei Teng}
\IEEEauthorblockA{\textit{Department of Electrical and Electronic Engineering, Imperial College London}, London, UK \\
\{chenghao.lyu26, ying.yu24, wangkun.xu18, f.teng\}@imperial.ac.uk}
}

\maketitle

\begin{abstract}
AI data centers are an emerging electricity demand, but limited local grid capacity could delay their connection and expansion. While existing operational strategies consider spatio-temporal workload flexibility, few studies address the strategic sizing and allocation of diverse accelerator archetypes with different capital and energy-efficiency trade-offs. This paper unlocks the flexibility potential of heterogeneous accelerators by jointly sizing and allocating different AI workloads (training, delay-tolerant batch work, and online inference requests) across sites and hours to meet the time-varying grid constraint  at minimum cost. Workload-specific throughput and energy coefficients link these decisions to facility power. In a representative two-site case, heterogeneous sizing reduces daily cost by 6.5\% relative to spatio-temporal scheduling with fixed capacity. When evaluated using workload-specific performance, a capacity design based on workload-averaged hardware coefficients costs 7.7\% more than the proposed design. The proposed design also eliminates the envelope violations of an unconstrained design at a 0.25\% daily cost increase. These results show how hardware heterogeneity and workload flexibility jointly shape economical, grid-compliant capacity plans.
\end{abstract}

\begin{IEEEkeywords}
AI data centers, capacity sizing, grid power envelopes, hardware heterogeneity, spatio-temporal workload scheduling.
\end{IEEEkeywords}

\section{Introduction}
AI data centers (AIDCs) are an emerging source of electricity demand, while limited local grid capacity can constrain their connection and expansion \cite{colangelo2026grid}. For an operator with multiple sites, a constrained grid connection does not necessarily require workload curtailment of AIDCs. Delay-tolerant work can be scheduled within its execution window, and online requests can be served at another site when the added network latency is acceptable. These options depend on the accelerator capacity installed at each site. To avoid operational bottlenecks during constrained grid hours or over-provisioning expensive hardware, accelerator capacity sizing and workload allocation should therefore be considered together.

Prior work has established the value of geographical workload shifting for renewable-energy utilization and data-center efficiency \cite{hao2023spatial}. More recent grid studies represent data-center demand as spatially shiftable or temporally deferrable load in power-system scheduling and expansion models \cite{radovanovic2023carbon, chen2026defer}. For online AI services, cross-site request routing has also been coordinated under grid power constraints while accounting for model deployment \cite{du2026model}. While these operational strategies effectively demonstrate the value of spatio-temporal workload flexibility, most of them dispatch tasks over fixed or homogeneous accelerators. Little research considers the strategic sizing and workload allocation across different accelerator archetypes that exhibit diverse capital expenditure and energy-efficiency trade-offs. Conversely, computing-serving studies account for these workload–hardware differences, but do not jointly consider hardware capacity sizing and workload allocation across sites under power system constraints \cite{jiang2025heterogeneous}.

Bridging this gap is important for unlocking AIDC flexibility. The same AI workload can exhibit different processing rates and energy consumption across different accelerator types. Additionally, training and offline batch workloads can be scheduled across sites and time, subject to their availability windows and completion deadlines \cite{chen2026defer}. In contrast, online inference requests must be served within their arrival hour, with additional costs incurred when requests are routed outside their originating regions \cite{du2026model}. The resulting optimization problem is to determine the number and types of accelerators deployed at each site, together with the spatial and temporal allocation of workloads, to minimize total cost while satisfying grid power constraints.

This paper presents an optimization model that jointly sizes heterogeneous accelerator capacity and schedules AI workloads. Training and offline tasks are placed across sites and hours within their deadlines. Online inference is served in its arrival hour, considering cross-site network delay penalties. Total facility power must comply with grid power envelopes, which are predefined hourly limits set to prevent local grid overload. The contributions include:
\begin{enumerate}
    \item \textit{Joint heterogeneous capacity sizing and workload allocation:} It formulates a multi-site optimization problem that jointly selects heterogeneous accelerator capacities, places delayable workload across sites and hours, and routes online inference requests under grid power limits.
    \item \textit{Hardware-aware facility power mapping:} Workload-specific throughput and energy coefficients link workload execution and active accelerator capacity to facility power. This formulation accounts for idle power, power usage effectiveness(PUE), and accelerator peak-power limits.
    \item \textit{Benchmark and deployment validation:} It constructs fixed-capacity and workload-averaged hardware baselines to isolate the value of heterogeneous capacity sizing. It then compares the continuous sizing solution with an 8-accelerator-per-server mixed-integer model to assess the modular deployment.
\end{enumerate}

\section{System Architecture and Hardware-Workload Taxonomy}

\subsection{Multi-Site Distributed AIDC Topology}
This paper considers a set of geographically distributed AIDC facilities, denoted by $d \in \mathcal{D}$, operated by a centralized entity over a discrete planning horizon $\mathcal{T} = \{1, \dots, T\}$ with uniform hourly resolution $\Delta t = 1.0\text{ h}$. Each AIDC $d$ connects with a regional grid substation characterized by time-varying electricity tariffs $\pi_{d,t}^{\mathrm{elec}}$ (\pounds/kWh), cooling infrastructure efficiency $\mathrm{PUE}_{d,t}$, and grid power envelopes $[P_{d,t}^{\mathrm{min}}, \bar{P}_{d,t}^{\mathrm{env}}]$\cite{petrou2021ensuring}. This work focuses on joint multi-site capacity sizing and spatio-temporal workload dispatch under predefined grid envelopes.

\subsection{AI Workload Taxonomy and Standardized Units}
To formulate diverse AI tasks within a unified optimization framework, computational demands are quantified in standardized work units, where one unit represents a nominal volume of processed data (tokens, samples, and queries). Task-specific variations in arithmetic complexity (FLOPs per token) and memory intensity are captured by empirical throughput coefficients $\mu_{g,k}$ (units/card/h) and dynamic energy parameters $e_{g,k}^{\mathrm{dyn}}$ (kWh/unit). The catalog encompasses three classes:
\begin{enumerate}
    \item \textit{Model Training ($w \in \mathcal{W}^{\mathrm{tr}}$):} Compute-intensive training require large parallel compute capacity ($W_w^{\mathrm{total}}$) across a window $[r_w, D_w]$, where $r_w$ is the arrival time and $D_w$ is the completion deadline.
    \item \textit{Delayable Offline Batch Workload ($w \in \mathcal{W}^{\mathrm{off}}$):} These comprise domain-specific model fine-tuning and offline inference. Let $\mathcal{W} = \mathcal{W}^{\mathrm{tr}} \cup \mathcal{W}^{\mathrm{off}}$ denote the complete set of delayable batch workloads. $A_w(t)$ denotes the cumulative processing data volume by hour $t$, and $\rho_{w,t}^{\mathrm{min}} \in [0, 1]$ is the benchmark cumulative progress fraction required.
    \item \textit{Online Inference Queries ($o \in \mathcal{O}$):} These are latency-critical queries (e.g., chat API endpoints) arriving exogenously from regional origin zones $r \in \mathcal{R}$ at time-varying request rates $\lambda_{o,r,t}$ (queries/h). Online requests cannot be deferred in time and must be dispatched to an available facility $d \in \mathcal{D}$ instantly. Disagreeing origin and execution locations ($d \ne r$) incur a quality-of-service (QoS) network latency penalty dictated by round-trip network delay $d_{r,d}^{\mathrm{lat}}$.
\end{enumerate}

\subsection{Heterogeneous Accelerators and Hardware-Task Compatibility}
To maintain generality and accommodate future hardware innovations, our framework models an arbitrary set of accelerator classes $g \in \mathcal{G}$ via a physical-economic parameter tuple: capital cost ($c_{d,g}^{\mathrm{cap}}$), peak power ($p_g^{\mathrm{peak}}$), idle static power ($p_g^{\mathrm{static}}$), task throughput ($\mu_{g,k}$), dynamic energy intensity ($e_{g,k}^{\mathrm{dyn}}$), and compatibility ($\kappa_{g,k}$). We instantiate two representative industry archetypes:
\begin{enumerate}
    \item \textit{Class H (High-Performance Compute-Dense Archetype):} High capital cost ($c_{d,\mathrm{H}}^{\mathrm{cap}}$), high peak power ($p_{\mathrm{H}}^{\mathrm{peak}}$), and notable idle draw ($p_{\mathrm{H}}^{\mathrm{static}}$), delivering superior absolute throughput on heavy pretraining ($\mu_{\mathrm{H},w}$).
    \item \textit{Class S (Cost- and Energy-Optimized Archetype):} Economical capital cost ($c_{d,\mathrm{S}}^{\mathrm{cap}}$), minimal idle leakage ($p_{\mathrm{S}}^{\mathrm{static}}$), and moderate peak power ($p_{\mathrm{S}}^{\mathrm{peak}}$), achieving high throughput-per-dollar and energy efficiency ($\mu_{\mathrm{S},k}/c_{d,\mathrm{S}}^{\mathrm{cap}}$, $e_{\mathrm{S},k}^{\mathrm{dyn}}$) for lightweight serving.
\end{enumerate}

This formulation couples computing demand with grid physics: the power parameters $(p_g^{\mathrm{static}}, e_{g,k}^{\mathrm{dyn}})$ determine substation demand, while $(\mu_{g,k}, \kappa_{g,k})$ dictate execution throughput and viability. The binary parameter $\kappa_{g,k} \in \{0, 1\}$ reflects architectural prerequisites (e.g., minimum precision). Any future accelerator technology can be evaluated within this unified model.

\section{Mathematical Problem Formulation}

\subsection{Optimization Objective Function}
It minimizes the total equivalent daily operational and amortized capital cost across the planning horizon $\mathcal{T}$:
\begin{equation}
\min C^{\mathrm{total}} = C^{\mathrm{cap}} + C^{\mathrm{elec}} + C^{\mathrm{lat}} + C^{\mathrm{unsrv}} + C^{\mathrm{unfin}} + C^{\mathrm{prog}}, \label{eq:obj}
\end{equation}
where each economic component is formulated as follows:
\begin{enumerate}
    \item \textit{Capital Cost ($C^{\mathrm{cap}}$):} Represents the daily capital expenditure (CAPEX) for accelerator capacities $C_{d,g} \ge 0$ across all sites:
\begin{equation}
C^{\mathrm{cap}} = \sum_{d, g} c_{d,g}^{\mathrm{cap}} C_{d,g}.
\end{equation}
    \item \textit{Grid Electricity Cost ($C^{\mathrm{elec}}$):} Captures the power purchasing bill dictated by time-varying wholesale tariffs $\pi_{d,t}^{\mathrm{elec}}$ and facility power $P_{d,t}^{\mathrm{fac}}$:
\begin{equation}
C^{\mathrm{elec}} = \sum_{d, t} \pi_{d,t}^{\mathrm{elec}} P_{d,t}^{\mathrm{fac}} \Delta t.
\end{equation}
    \item \textit{QoS Degradation Penalty ($C^{\mathrm{lat}}$):} Penalizes cross-site execution when online query traffic from origin region $r$ is routed to a distant data center $d \ne r$, proportional to round-trip latency $d_{r,d}^{\mathrm{lat}}$ and unit penalty $\alpha_o^{\mathrm{lat}}$:
\begin{equation}
C^{\mathrm{lat}} = \sum_{d, g, o, r, t} \alpha_o^{\mathrm{lat}} d_{r,d}^{\mathrm{lat}} x_{d,g,o,r,t}^{\mathrm{on}}.
\end{equation}
    \item \textit{Unserved Online Query Loss ($C^{\mathrm{unsrv}}$):} Imposes a reliability penalty $\alpha_o^{\mathrm{unsrv}}$ on any unfulfilled online queries $q_{o,r,t}^{\mathrm{unsrv}}$:
\begin{equation}
C^{\mathrm{unsrv}} = \sum_{o, r, t} \alpha_o^{\mathrm{unsrv}} q_{o,r,t}^{\mathrm{unsrv}}.
\end{equation}
    \item \textit{Unfinished Workload Penalty ($C^{\mathrm{unfin}}$):} Evaluated at deadline $D_w$, assessing financial penalties $c_w^{\mathrm{unfin}}$ on any residual unfinished work units $q_w^{\mathrm{unfin}}$:
    \begin{equation}
    C^{\mathrm{unfin}} = \sum_{w \in \mathcal{W}} c_w^{\mathrm{unfin}} q_w^{\mathrm{unfin}}.
    \end{equation}
    \item \textit{Intermediate Progress Deficit Penalty ($C^{\mathrm{prog}}$):} Penalizes cumulative progress shortfalls $v_{w,t}^{\mathrm{prog}}$ relative to target milestones $\rho_{w,t}^{\mathrm{min}}$ at hourly rate $c_w^{\mathrm{prog}}$, discouraging bursty deferrals:
\begin{equation}
C^{\mathrm{prog}} = \sum_{w, t \le D_w} c_w^{\mathrm{prog}} v_{w,t}^{\mathrm{prog}} \Delta t.
\end{equation}
\end{enumerate}

\subsection{System Operational and Grid Coupling Constraints}
The operational space is bounded by seven physical and coupling constraint groups:

1) \textit{Accelerator Allocation:} At each operational hour $t$, the total number of active accelerators dedicated to offline (training and batch) jobs ($u_{d,g,w,t}^{\mathrm{off}} \ge 0$) and online queries ($u_{d,g,o,r,t}^{\mathrm{on}} \ge 0$) cannot exceed capacity:
\begin{equation}
\sum_{w \in \mathcal{W}} u_{d,g,w,t}^{\mathrm{off}} + \sum_{o \in \mathcal{O}, r \in \mathcal{R}} u_{d,g,o,r,t}^{\mathrm{on}} \le C_{d,g},  \forall d \in \mathcal{D}, g \in \mathcal{G}, t \in \mathcal{T}. \label{eq:cap_alloc}
\end{equation}

2) \textit{Hardware Compatibility and Throughput Conversion:} Following established AI cluster scheduling frameworks and heterogeneous serving systems, workload execution volume is bounded by active allocated capacity $u$ and task-specific productivity $\mu_{g,k}$. A binary compatibility indicator $\kappa_{g,k} \in \{0, 1\}$ strictly prohibits assignment when hardware lacks required precision or memory capacity:
\begin{align}
x_{d,g,w,t}^{\mathrm{off}} &\le \kappa_{g,w} \mu_{g,w} u_{d,g,w,t}^{\mathrm{off}} \Delta t, \quad \forall d, g, w, t, \label{eq:through_batch} \\
x_{d,g,o,r,t}^{\mathrm{on}} &\le \kappa_{g,o} \mu_{g,o} u_{d,g,o,r,t}^{\mathrm{on}} \Delta t, \quad \forall d, g, o, r, t. \label{eq:through_online}
\end{align}

3) \textit{Arrival, Deadline Completion, and Milestones:} Workload execution must respect causality, deadlines, and intermediate execution schedules:
\begin{align}
\sum_{\tau=1}^t \sum_{d \in \mathcal{D}} \sum_{g \in \mathcal{G}} x_{d,g,w,\tau}^{\mathrm{off}} &\le A_w(t), \quad \forall w \in \mathcal{W}, t \in \mathcal{T}, \label{eq:causal_arr} \\
\sum_{t \le D_w} \sum_{d \in \mathcal{D}} \sum_{g \in \mathcal{G}} x_{d,g,w,t}^{\mathrm{off}} + q_w^{\mathrm{unfin}} &= W_w^{\mathrm{total}}, \quad \forall w \in \mathcal{W}, \label{eq:deadline_comp} \\
\sum_{\tau=1}^t \sum_{d \in \mathcal{D}} \sum_{g \in \mathcal{G}} x_{d,g,w,\tau}^{\mathrm{off}} + v_{w,t}^{\mathrm{prog}} &\ge \rho_{w,t}^{\mathrm{min}} W_w^{\mathrm{total}}, \quad \forall w, t \le D_w, \label{eq:progress_ms}
\end{align}
alongside $x_{d,g,w,t}^{\mathrm{off}} = 0$ for all $t > D_w$. Constraint \eqref{eq:causal_arr} ensures the workload cannot be processed prior to arrival. Constraint \eqref{eq:deadline_comp} requires total completed work plus penalty slack $q_w^{\mathrm{unfin}}$ to clear total demand $W_w^{\mathrm{total}}$ by deadline $D_w$. In \eqref{eq:progress_ms}, milestone curve $\rho_{w,t}^{\mathrm{min}}$ enforces steady computational progress.

4) \textit{Online Query Conservation:} Real-time queries $\lambda_{o,r,t}\Delta t$ arriving from origin $r$ must either be processed across available data centers or registered as unserved loss:
\begin{equation}
\sum_{d \in \mathcal{D}} \sum_{g \in \mathcal{G}} x_{d,g,o,r,t}^{\mathrm{on}} + q_{o,r,t}^{\mathrm{unsrv}} = \lambda_{o,r,t}\Delta t, \quad \forall o \in \mathcal{O}, r \in \mathcal{R}, t \in \mathcal{T}. \label{eq:online_flow}
\end{equation}

5) \textit{Dynamic Power Cap:} Although dynamic energy scales with processed workload via $e_{g,k}^{\mathrm{dyn}}$, hardware possesses a thermal design power ($p_g^{\mathrm{peak}}$). Because idle power accounts for $p_g^{\mathrm{static}}$, the remaining dynamic power headroom is bounded by:
\begin{equation}
\frac{1}{\Delta t} \sum_{k \in \mathcal{K}} e_{g,k}^{\mathrm{dyn}} x_{d,g,k,t} \le (p_g^{\mathrm{peak}} - p_g^{\mathrm{static}}) u_{d,g,t}^{\mathrm{act}}, \quad \forall d, g, t, \label{eq:peak_power}
\end{equation}
where $u_{d,g,t}^{\mathrm{act}} \triangleq \sum_{w \in \mathcal{W}} u_{d,g,w,t}^{\mathrm{off}} + \sum_{o \in \mathcal{O}, r \in \mathcal{R}} u_{d,g,o,r,t}^{\mathrm{on}}$.

6) \textit{Facility Power Mapping:} Total facility power encompasses IT computation and temperature-dependent cooling overhead. Total IT energy $E_{d,t}^{\mathrm{IT}}$ includes fixed base power $P_d^{\mathrm{IT,fix}}$, idle power of allocated cards, and dynamic workload computing energy:
\begin{align}
E_{d,t}^{\mathrm{IT}} &= P_d^{\mathrm{IT,fix}} \Delta t + \sum_{g \in \mathcal{G}} p_g^{\mathrm{static}} u_{d,g,t}^{\mathrm{act}} \Delta t \nonumber \\
&\quad + \sum_{g,w} e_{g,w}^{\mathrm{dyn}} x_{d,g,w,t}^{\mathrm{off}} + \sum_{g,o,r} e_{g,o}^{\mathrm{dyn}} x_{d,g,o,r,t}^{\mathrm{on}}. \label{eq:it_energy}
\end{align}
Total facility electrical power $P_{d,t}^{\mathrm{fac}}$ is then calculated as:
\begin{equation}
P_{d,t}^{\mathrm{fac}} = P_d^{\mathrm{fac,fix}} + \mathrm{PUE}_{d,t} \cdot \left(\frac{E_{d,t}^{\mathrm{IT}}}{\Delta t}\right), \quad \forall d \in \mathcal{D}, t \in \mathcal{T}, \label{eq:pue_power}
\end{equation}
where $P_d^{\mathrm{fac,fix}}$ is non-IT facility power. $\mathrm{PUE}_{d,t} \ge 1.0$ is predefined based on ambient temperature.

7) \textit{Grid Power Envelope:} To ensure grid safety, total facility power must respect regional network boundaries:
\begin{equation}
P_{d,t}^{\mathrm{min}} \le P_{d,t}^{\mathrm{fac}} \le \bar{P}_{d,t}^{\mathrm{env}}, \quad \forall d \in \mathcal{D}, t \in \mathcal{T}. \label{eq:grid_envelope}
\end{equation}

\subsection{Baseline Formulation and Nested Hierarchy}
To rigorously isolate individual flexibility mechanisms, four comparison baselines are formulated under unified penalty structures and fixed baseline capacity $C_{d,g}^{\mathrm{base}}$:
\begin{itemize}
    \item \textbf{Model A (Rigid Reference Schedule):} Executes workloads strictly against a non-flexible reference schedule $\bar{x}_{w,t}^{\mathrm{ref}}$ fixed to hourly arrivals at designated origin site $d_w^0$:
    \begin{equation}
    \sum_{g \in \mathcal{G}} x_{d_w^0,g,w,t}^{\mathrm{off}} + z_{w,t}^{\mathrm{rigid}} = \bar{x}_{w,t}^{\mathrm{ref}}, \quad x_{d,g,w,t}^{\mathrm{off}} = 0 \; (\forall d \ne d_w^0), \label{eq:rigid_sched}
    \end{equation}
    with deficits $z_{w,t}^{\mathrm{rigid}}$ penalized at rate $c_w^{\mathrm{unfin}}$, and cross-site routing prohibited.
    \item \textbf{Model B (Local Temporal Shifting):} Relaxes rigid tracking to cumulative causal bounds \eqref{eq:causal_arr}--\eqref{eq:progress_ms}, allowing temporal shifting within $[r_w, D_w]$ without spatial routing ($x_{d,g,w,t}^{\mathrm{off}} = 0, \forall d \ne d_w^0$). Since any schedule satisfying \eqref{eq:rigid_sched} satisfies \eqref{eq:causal_arr}--\eqref{eq:progress_ms}, the feasible set satisfies $\mathcal{F}_A \subseteq \mathcal{F}_B$, mathematically guaranteeing $C_B \le C_A$.
    \item \textbf{Model C (Spatio-Temporal Dispatch):} Enables inter-site workload routing across facilities under fixed capacity $C_{d,g}^{\mathrm{base}}$. Since $\mathcal{F}_B \subseteq \mathcal{F}_C$, it strictly holds that $C_C \le C_B$.
    \item \textbf{Model D (Proposed Joint Co-Optimization):} Simultaneously co-optimizes continuous capacity $C_{d,g}$ and hourly spatio-temporal dispatch under grid envelopes \eqref{eq:grid_envelope}, expanding the optimization domain over fixed $C_{d,g}^{\mathrm{base}}$ and guaranteeing $C_D \le C_C$.
    \item \textbf{8-Card Server MILP:} Enforces discrete rack procurement modularity on Model D:
    \begin{equation}
    C_{d,g} = 8 \cdot y_{d,g}, \quad y_{d,g} \in \mathbb{Z}^+, \quad \forall d \in \mathcal{D}, g \in \mathcal{G}. \label{eq:milp_sizing}
    \end{equation}
   \item \textbf{Model E (Workload-Agnostic Baseline):} Sizes capacity using workload-averaged throughput and energy:
\begin{equation}
\bar{\mu}_g = \sum_{k} \omega_k \mu_{g,k}, \quad \bar{e}_g^{\mathrm{dyn}} = \sum_{k} \omega_k e_{g,k}^{\mathrm{dyn}}, \quad \forall g \in \mathcal{G}, \label{eq:model_e_agg}
\end{equation}
where $\omega_k$ is the share of workload $k$. The resulting capacity is then tested under actual demands to evaluate the cost of ignoring workload differences.
\end{itemize}

\section{A Representative Dual-Site Case Study}

\subsection{Benchmark Parameterization and System Setup}
The proposed methodology is validated on a representative dual-site benchmark inspired by the UK transmission network, comprising Site 1 in London ($d = \text{London}$, Southern UK origin $r = \text{South\_UK}$) and Site 2 in Edinburgh ($d = \text{Edinburgh}$, Northern UK origin $r = \text{North\_UK}$). London represents an urban load center with dense interactive demand, high daytime electricity tariffs (\pounds 0.22--\pounds 0.32/kWh), tight grid power limits ($\bar{P}_{\text{Lon},t}^{\mathrm{env}} = 125\text{ kW}$ peak), and $\mathrm{PUE} = 1.28$. Edinburgh represents with abundant wind generation, lower electricity tariffs \cite{desnz2023dukes} (\pounds 0.18--\pounds 0.26/kWh), higher grid capacity ($\bar{P}_{\text{Edi},t}^{\mathrm{env}} = 185\text{ kW}$), and cooler climate ($\mathrm{PUE} = 1.20$), where temperature-dependent PUE profiles are calibrated against ASHRAE thermal guidelines \cite{ashrae2021thermal}. The inter-site network round-trip delay is set to 14~ms based on UK optical backbone measurements. Accelerator Class H is calibrated to state-of-the-art 700~W compute-dense training GPUs (e.g., NVIDIA H100 SXM, $p_H^{\mathrm{peak}} = 0.70\text{ kW}, p_H^{\mathrm{static}} = 0.12\text{ kW}, c_H^{\mathrm{cap}} = \text{\pounds}26/\text{card/day}$ amortized over a standard 3-year depreciation cycle \cite{jiang2025heterogeneous, nvidia2023h100}), while Class S represents power-optimized inference accelerators (e.g., 350--400~W enterprise GPUs, $p_S^{\mathrm{peak}} = 0.40\text{ kW}, p_S^{\mathrm{static}} = 0.05\text{ kW}, c_S^{\mathrm{cap}} = \text{\pounds}8.5/\text{card/day}$).

\subsection{Benchmark Comparison}
The optimization models were solved using Gurobi 12.0 under Python 3.9 on an Apple M2 chip. All continuous LP models solved to exact optimality in under 0.15 seconds, while the 8-card MILP solved in 0.33 seconds. Table~\ref{tab:model_comparison} reports the cost breakdown.

\begin{table}[t]
\centering
\caption{Equivalent Daily Cost Breakdown}
\label{tab:model_comparison}
\renewcommand{\arraystretch}{0.92}
\resizebox{\columnwidth}{!}{
\begin{tabular}{lccccc}
\toprule
\textbf{Model} & \textbf{Total (\pounds)} & \textbf{CAPEX (\pounds)} & \textbf{Elec (\pounds)} & \textbf{QoS (\pounds)} & \textbf{Penalty (\pounds)} \\
\midrule
Model A (Rigid)      & \pounds 25,060.8 & \pounds 4,037.5 & \pounds 789.0 & \pounds 245.0 & \pounds 19,989.2 \\
Model B (Temporal)   & \pounds 25,020.4 & \pounds 4,037.5 & \pounds 775.9 & \pounds 245.0 & \pounds 19,961.9 \\
Model C (Spatio-Temp)& \pounds 5,653.8  & \pounds 4,037.5 & \pounds 862.6 & \pounds 331.0 & \pounds 422.7 \\
\textbf{Model D (Proposed)} & \textbf{\pounds 5,288.6} & \textbf{\pounds 4,120.2} & \textbf{\pounds 864.5} & \textbf{\pounds 303.9} & \textbf{\pounds 0.0} \\
Model E (Workload-Agn.)  & \pounds 5,773.5  & \pounds 4,511.5 & \pounds 976.2 & \pounds 285.8 & \pounds 0.0 \\
\bottomrule
\end{tabular}
}
\end{table}

\begin{enumerate}
    \item \textit{Value of Temporal Shifting (A $\to$ B):} Shifting delayable batch workload within release and deadline windows provides a saving of \pounds 40.34/day. However, severe local capacity bottlenecks in London still result in \pounds 9,948.75/day in unserved queries.
    \item \textit{Value of Spatial Routing (B $\to$ C):} Enabling inter-site routing reduces total daily costs by 77.4\%. Spatially routing congested Southern queries to Edinburgh avoids the unserved penalty.
    \item \textit{Value of Heterogeneous Capacity Sizing (C $\to$ D):} The proposed Model D optimizes accelerator sizing, allocating 30.0 Class H and 43.35 Class S in London, alongside 77.05 Class H and 154.81 Class S in Edinburgh. This saves an additional \pounds 365.16/day over fixed baseline sizing.
\end{enumerate}

\subsection{Operating Envelope Compliance}
\begin{figure}[t]
\centering
\includegraphics[width=0.88\columnwidth]{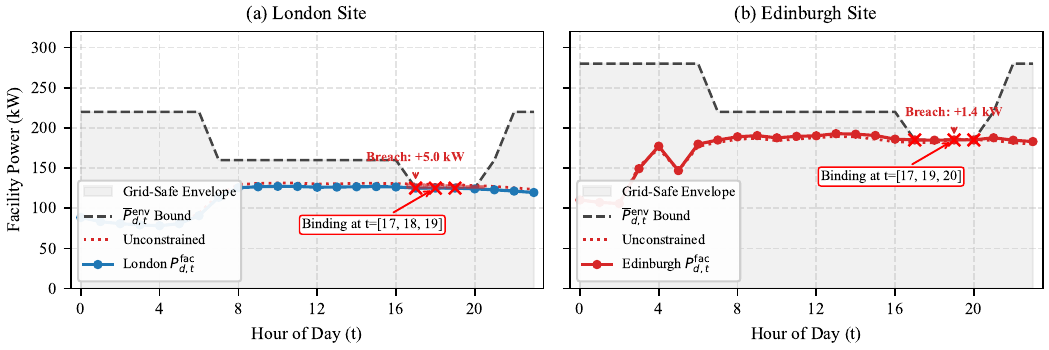}
\caption{Facility power with grid operating envelopes under Model D.}
\label{fig:grid_power}
\end{figure}

\begin{figure}[t]
\centering
\includegraphics[width=0.88\columnwidth]{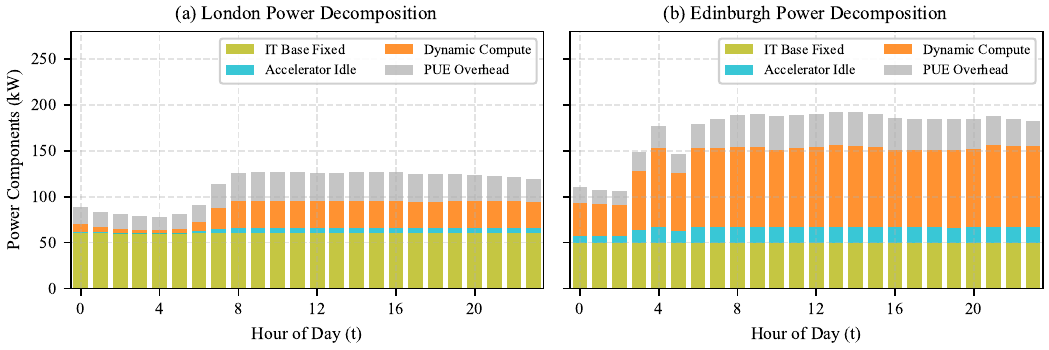}
\caption{Facility power composition breakdown.}
\label{fig:power_decomp}
\end{figure}

Fig.~\ref{fig:grid_power} illustrates facility power tracking against gridoperating envelopes. In Model D, London facility power hits the 125~kW upper bound during hours $t \in [17, 18, 19]$, and Edinburgh binds at 185~kW during hours $t \in [17, 19, 20]$. To quantify the economic impact of grid compliance, an unconstrained model was solved. As shown in Fig.~\ref{fig:counterfactual}(a), the London facility power surges to 130.04~kW during peak hours, violating gridsafety limits across four consecutive hours with a peak exceedance of 5.04~kW. The proposed method avoids this grid risk at a marginal shadow cost of only \pounds 13.17/day. Fig.~\ref{fig:power_decomp} details the physical decomposition across fixed IT, idle draw, dynamic load, and cooling overhead.

\subsection{Location-Aware Online Inference Routing}
\begin{figure}[t]
\centering
\includegraphics[width=0.88\columnwidth]{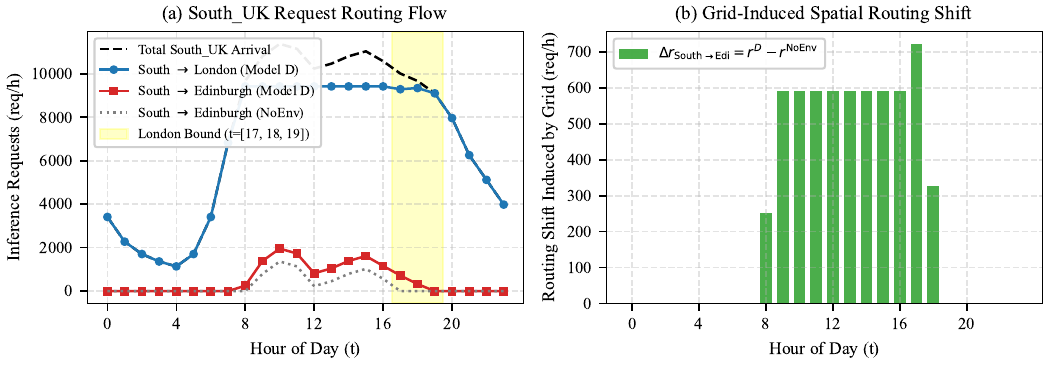}
\caption{Location-aware online inference routing: (a) origin-destination dispatch; (b) differential cross-site routing proving peak-hour congestion diversion.}
\label{fig:online_routing}
\end{figure}

Fig.~\ref{fig:online_routing} details the location-aware spatial routing of ChatAPI queries under joint latency and grid constraints. In normal hours, Southern UK queries are served locally in London to minimize network latency QoS penalties. However, during the evening grid congestion window, the differential routing metric $\Delta r_{\text{South} \to \text{Edi}}(t) = r_t^D - r_t^{\text{NoEnv}}$ (Fig.~\ref{fig:online_routing}(b)) shows that Model D reroutes 7.29\% of Southern queries to Edinburgh. This spatial diversion provides peak shaving without curtailing user requests.

\subsection{Practical Validity of Continuous LP under Modular Server}
In real-world deployments, AI accelerators are procured as standardized modular server nodes (e.g., 8-card chassis, $C_{d,g} = 8 y_{d,g}, y_{d,g} \in \mathbb{Z}^+$). To address this, we solved the exact 8-card node MILP and decomposed the deployment cost in Table~\ref{tab:milp_audit}. On the domain-aligned feasible space ($C_{\text{Lon},H} \le 24$ cards, 3 nodes), the exact MILP obtains \pounds 5,362.90/day compared to the continuous LP of \pounds 5,330.11/day. Therefore, the pure mathematical integrality gap is 0.62\%. The remaining 0.78\% (\pounds 41.48/day) is from modular discretization loss.

\begin{table}[t]
\centering
\caption{Exact MILP and Continuous Relaxation Decomposition}
\label{tab:milp_audit}
\renewcommand{\arraystretch}{0.92}
\begin{tabular}{lc}
\toprule
\textbf{Configuration}& \textbf{Daily Cost (\pounds/d)} \\
\midrule
Ideal Continuous LP ($C_{\text{Lon},H} \le 30$)   & \pounds 5,288.63 \\
Aligned Continuous LP ($C_{\text{Lon},H} \le 24$) & \pounds 5,330.11 \\
Exact 8-Card Node MILP ($y_{d,g} \in \mathbb{Z}^+$) & \pounds 5,362.90 \\
\midrule
\textbf{Mathematical Integrality Gap}& \textbf{+\pounds 32.79 (+0.62\%)} \\
\textbf{Modular Loss}& \textbf{+\pounds 41.48 (+0.78\%)} \\
\midrule
\textbf{Practical Deployment}       & \textbf{+\pounds 74.27 (+1.40\%)} \\
\bottomrule
\end{tabular}
\end{table}

\subsection{Workload-Agnostic Average Baseline (Model E)}
Model E optimizes capacity assuming workload-weighted average processing throughput ($\bar{\mu}_H=183.2, \bar{\mu}_S=73.5$) and dynamic energy ($\bar{e}_H^{\mathrm{dyn}}=0.003025, \bar{e}_S^{\mathrm{dyn}}=0.004740\text{ kWh/unit}$). Model E obtains a cost of \pounds 5,773.49/day without penalties. However, since average metrics hide the strict necessity of Class H compute density for LLM pretraining, Model E overinvests in Class S accelerators. This misallocation increases daily CAPEX to £4,511.50/day, which is 9.50\% higher than Model D. It proves the priority of heterogeneous accelerator sizing.

\subsection{Grid Power Envelope Tightening and Dynamic Load Response}
While baseline envelope compliance introduces a modest shadow cost (+0.25\%), the core resilience of spatio-temporal co-optimization lies in extreme urban grid bottlenecks under progressive capacity restrictions. To evaluate this, the envelopes were tightened by $\eta \in \{0\%, 10\%, 20\%, 30\%, 40\%\}$ during the evening peak time. 

\begin{figure}[t]
\centering
\includegraphics[width=0.88\columnwidth]{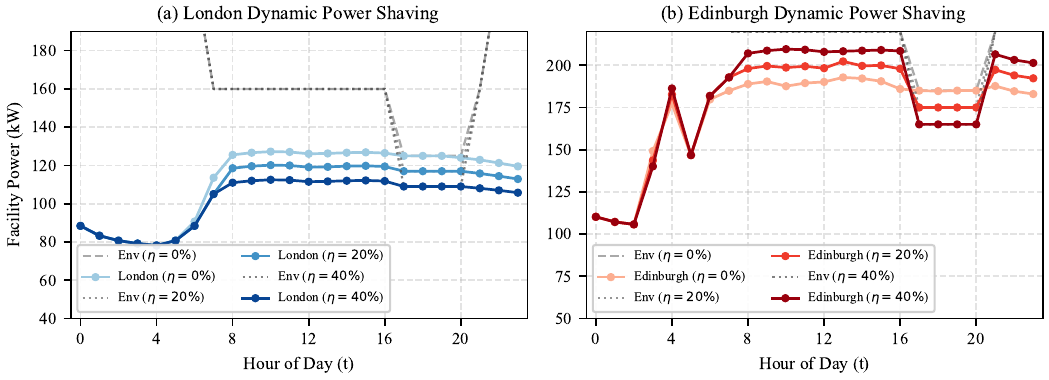}
\caption{Facility power under grid envelope tightening.}
\label{fig:counterfactual}
\end{figure}

As shown in Fig.~\ref{fig:counterfactual}, the facility power curves exhibit a step-wise peak shaving response to contracting grid boundaries. In London (Fig.~\ref{fig:counterfactual}(a)), as the envelope power is compressed from 125~kW down to 109~kW, the facility power curve follows the envelope. Meanwhile, Edinburgh (Fig.~\ref{fig:counterfactual}(b)) dynamically coordinates its schedule to absorb the geographically diverted workloads within its local envelope. 

The economic and routing trade-offs driving this adaptation are summarized in Table~\ref{tab:sensitivity}. As $\eta$ tightens from 0\% to 40\%, total binding hours expand from 6 to 8 hours, and the system dynamically increases Southern cross-site diversion from 7.29\% to 23.80\%. This proactive spatial shifting permits local substations to operate safely while restricting overall daily cost growth to only 3.91\%.

\begin{table}[t]
\centering
\caption{Grid Envelope Tightening Sensitivity Analysis}
\label{tab:sensitivity}
\renewcommand{\arraystretch}{0.92}
\resizebox{\columnwidth}{!}{
\begin{tabular}{cccccc}
\toprule
$\boldsymbol{\eta}$ \textbf{(\%)} & \textbf{Cost (\pounds/d)} & \textbf{CAPEX (\pounds)} & \textbf{Elec (\pounds)} & \textbf{Binding} & \textbf{Cross-Site} \\
\midrule
0\%  & \pounds 5,288.6 & \pounds 4,120.2 & \pounds 864.5 & 6 h & 7.3\% \\
10\% & \pounds 5,337.8 & \pounds 4,156.8 & \pounds 859.7 & 8 h & 10.4\% \\
20\% & \pounds 5,388.5 & \pounds 4,190.7 & \pounds 853.9 & 8 h & 14.4\% \\
30\% & \pounds 5,439.9 & \pounds 4,224.8 & \pounds 848.2 & 8 h & 18.5\% \\
40\% & \pounds 5,495.5 & \pounds 4,257.2 & \pounds 841.8 & 8 h & 23.8\% \\
\bottomrule
\end{tabular}
}
\end{table}

\section{Conclusion}
This paper unlocks the flexibility potential of heterogeneous accelerators by presenting a framework that jointly sizes heterogeneous capacity and allocates AI workload across sites and hours under predefined grid power envelopes. Workload-specific throughput and energy coefficients connect hardware allocation and workload to facility power. In a representative two-site case, joint sizing and scheduling reduced equivalent daily cost by 6.5\% relative to spatio-temporal scheduling with fixed capacity. A design based on workload-averaged hardware coefficients incurred a 7.7\% higher realized cost when evaluated using workload-specific performance. Enforcing the grid power envelopes eliminated the violations of an unconstrained design at a 0.25\% increase in daily cost. A comparison with an eight-accelerator-per-server mixed-integer formulation also quantified the cost of modular deployment. These results indicate that grid-compliant capacity plans depend on matching accelerator types to workload requirements and using the distinct spatial and temporal flexibility of each workload type.

The model addresses initial allocation and hourly request routing, not the runtime migration of training states. Incorporating uncertain demand and grid limits remains future work.

\bibliographystyle{IEEEtran}
\bibliography{references}

\end{document}